\documentclass[12pt]{article}

\usepackage{epsf}
\usepackage{epsfig}
\usepackage{cite}
\usepackage{amsmath}
\usepackage{graphics}

\begin{document}

\setlength{\textheight}{21.5cm}
\setlength{\oddsidemargin}{0.cm}
\setlength{\evensidemargin}{0.cm}
\setlength{\topmargin}{0.cm}
\setlength{\footskip}{1cm}
\setlength{\arraycolsep}{2pt}

\renewcommand{\thefootnote}{\#\arabic{footnote}}
\setcounter{footnote}{0}

\newcommand{\gtrsim}{ \mathop{}_{\textstyle \sim}^{\textstyle >} }
\newcommand{\lesssim}{ \mathop{}_{\textstyle \sim}^{\textstyle <} }
\newcommand{\rem}[1]{{\bf #1}}
\renewcommand{\thefootnote}{\fnsymbol{footnote}}
\setcounter{footnote}{0}
\def\thefootnote{\fnsymbol{footnote}}

\hfill {\tt arXiv:08mm.nnnn[hep-ph]}\\
\vskip .5in

\begin{center}

\bigskip
\bigskip

{\Large \bf Doubly-Charged Boson 
Production in $p \bar{p}$ Collisions at $\sqrt{s} = 1.96$ TeV}

\vskip .45in

{\bf Paul H. Frampton\footnote{frampton@physics.unc.edu}} 

\vskip .3in

{\it Department of Physics and Astronomy, University of North Carolina,
Chapel Hill, NC 27599-3255.}

\end{center}

\vskip .4in 
\begin{abstract}
It is suggested that the observed excess
of muons in studies of $p \bar{p}$ collisions 
at $\sqrt{s} = 1.96$ TeV arises from the decay
of doubly-charged bosons. 
Such particles are predicted in the 3-3-1 model 
where the electroweak gauge group
is extended to $SU(3) \times U(1)$.

\end{abstract}

\renewcommand{\thepage}{\arabic{page}}
\setcounter{page}{1}
\renewcommand{\thefootnote}{\#\arabic{footnote}}

\newpage

\bigskip

\noindent Knowledge from particle experiments of
physics up to 100 GeV allow us to extrapolate
back to a time about $10^{-10}$ second after the Big
Bang with confidence. To go further back
in the early universe to times $10^{-12}$ second
after the Big Bang and earlier, which is crucial
to our understanding of early
cosmology, awaits clarification
of particle physics up to TeV scales and beyond.

\bigskip

\noindent After many confirmations and reconfirmations of
the standard model of particle phenomenology
at scales up to a few 100 GeV, ones is awaiting
TeV energy scales, especially at the Large Hadron
Collider (LHC) for data on the Higgs boson
and on new physics.

\bigskip

\noindent Many suggestions have been made of how the
standard model of particle phenomenology extends
to higher energies in the TeV range. Since the discovery
of the $W^{\pm}$ and $Z$ gauge bosons in the 1980's
played a key role is confirming the gauge structure
it is natural to suggest that there are more
gauge bosons at the TeV scale which reflect
an enlargement of the gauge group. There will
typically also be more scalars.

\bigskip

\noindent There have been many apparent anomalies 
in experiment suggesting disagreement with the standard
model, ambulances to chase, and with the notable
exception of neutrino
oscillations these have
disappeared as more data emerged. So one must be
cautious about seizing on any new anomaly before
waiting confirmation of the data.

\bigskip

\noindent Nevertheless, in the present note we offer 
a possible interpretation
of the large excess of muons recently reported\cite{CDF1}
by the CDF group at Fermilab in $p\bar{p}$
collisions at $1.96$ TeV.

\bigskip

\noindent The additional muons have an impact
parameter $\sim 1$ cm relative to the collision
suggesting a possible lifetime $\sim$ 10 ps.
The cross-section \cite{CDF1,CDF2} for their production
is $\sim 1$ nb. A phenomenological model
was suggested in \cite{CDF3}.

\bigskip

\noindent In the pair production of muons there is a surprising
number of same sign (SS) charges as opposite sign (OS)
charges. There are 10,00's such events and the ratio
$(SS)/(OS) \sim 1/2$. Most mechanisms will produces
far more OS pairs than suggested by this ratio.

\bigskip

\noindent Nevertheless, this ratio of same versus
opposite sign is expected if the primorial process
is pair producing doubly-charged particles as
$X^{++}X^{--}$. The resultant charges have
$(SS)/(OS) \simeq 1/2$.

\bigskip

\noindent In \cite{CDF3} new states are hypothesized
with masses up to 15 GeV. However, it is equally possible
with the data available that much
heavier states in the TeV mass range are being
produced as will be discussed here.

\bigskip

\noindent Doubly-charged gauge bosons and scalars
in the TeV range are predicted by 
the 3-3-1 model \cite{331,3312}. In the 3-3-1
model three families are required by cancellation
of triangle anomalies between families, the third
family involving the top quark being treated
asymmetrically with respect to the first
two families
\footnote{Note that the
phenomenologically allowed
choice of family to be treated
asymmetrically in the 3-3-1 model
is the third one, as in \cite{331},
not the first one as in \cite{3312}.}.

\bigskip

\noindent Enlargement of the electroweak
gauge group to $SU(3) \times U(1)$
leads to five additional gauge bosons.
They are a $Z^{'}$ and four bileptons
$(Y^{++}, Y^{+})$, $(Y^{--}, Y^{-})$.
From low-energy experiments, particularly
''wrong" (V+A) muon decay and
muonium-antimuonium conversion,
the lower bound on the bilepton
mass is about 1 TeV \cite{CD}.

\bigskip

\noindent The production cross-section
for bileptons in hadron collisions is in
\cite{CD,FLR}. The production and
immediate decay of bileptons into muons cannot
explain the data because the observed
impact parameters\cite{CDF1} correspond
to too long a bilepton decay lifetime.

\bigskip

\noindent The bilepton decay width for $Y^{\pm\pm} \rightarrow
\mu^{\pm}\mu^{\pm}$ can be estimated
as $\Gamma\sim \alpha_2 M_Y$ where $\alpha_2$
is the weak coupling and $M_Y \sim 1$ TeV.
This leads to a lifetime $\tau \le 10^{-24}$ sec.
which is many orders
of magnitude smaller than needed to explain
the data which require $\tau \ge 10^{-12}$ s.
At first sight then, such an interpretation
as bilepton production looks hopeless for
this reason alone.

\bigskip

\noindent Upon further reflection, however,
one notices in \cite{331} that the symmetry
breaking to the standard electroweak group
necessarily involves scalars which are
double charged in triplets 
$(T^{\pm\pm}, T^{\pm}, T^0)$ of which
the vacuum value $<T^0>$ breaks the
symmetry. The doubly charged scalars
couple much more weakly to muons
by Yukawa couplings $\sim m_{\mu}/M_W \simeq 
10^{-6}$ so cascading of
a bilepton into scalars could be consistent
with the particle longevity implied by the data.

\bigskip

\noindent Primordial appearance of $T^{++}$
in the $p\bar{p}$ collisions is unlikely to account
for the observed muon excess because, for the same
reason that the $T$ lifetime is longer,
the production cross-section is too small.
Direct production
of bileptons which decay $Y^{++} \rightarrow
T^{++} + .... \rightarrow \mu^{+}\mu^{+}$
seems a likely possibility.
With the present available data,
estimation of the
masses for the bilepton $Y$ and the
scalar $T$ is not yet possible. It
is worth careful 
experimental study of the bimuon mass distributions
viewed as low-energy tails of TeV-scale
doubly-charged boson decays.

\bigskip

\noindent It would be interesting if 
such further analysis of the CDF data in \cite{CDF1}
reveal that it 
has scooped LHC, not
for the discovery of the standard Higgs
rather for the first signal
of an extended gauge sector and
bileptons $Y^{\pm\pm}, Y^{\pm}$
heavier than $W^{\pm}$ and $Z$.

\bigskip

\newpage

\begin{center}

\section*{Acknowledgements}

\end{center}

\noindent I thank M. Kruse for discussion of the recent
CDF data. This work was supported in part 
by the U.S. Department of Energy under Grant
No. DE-FG02-06ER41418.

\bigskip
\bigskip
\bigskip
\bigskip

\end{document}